\documentclass[twocolumn,prb,preprintnumbers,amsmath,superscriptaddress,amssymb]{revtex4-1}
\usepackage[section]{placeins}
\usepackage{graphicx}
\usepackage{amsmath,amssymb}
\usepackage{color}
\usepackage{multirow}
\usepackage[latin1]{inputenc}
\usepackage[T1]{fontenc}
\usepackage[english]{babel}
\usepackage{amsmath,amssymb,amsthm}
\usepackage{hyperref}
\usepackage{booktabs}
\usepackage{bbm}
\usepackage{url}
\usepackage{nicefrac}
\usepackage{multirow} 
\usepackage{graphicx}
\usepackage{array}
\usepackage{times,txfonts}
\usepackage{array}
\renewcommand{\epsilon}{\varepsilon}
\renewcommand{\phi}{\varphi}

\newcolumntype{L}[1]{>{\raggedright\let\newline\\\arraybackslash\hspace{0pt}}m{#1}}
\newcolumntype{C}[1]{>{\centering\let\newline\\\arraybackslash\hspace{0pt}}m{#1}}
\newcolumntype{R}[1]{>{\raggedleft\let\newline\\\arraybackslash\hspace{0pt}}m{#1}}
\usepackage{sistyle}
\usepackage{soul}
\definecolor{lightblue}{RGB}{185,210,248}
\sethlcolor{lightblue}

\begin{document}

\title{Secure communication using low dimensional topological elements}
\author{Manuel F. Ferrer-Garcia}
\affiliation{Nexus for Quantum Technologies, University of Ottawa, Ottawa, K1N 6N5, ON, Canada}
\author{Avishy Carmi}
\affiliation{Center for Quantum Information Science and Technology and the Faculty of Engineering Sciences, Ben-Gurion University of Negev, Beersheba, Israel}
%%%
\author{Alessio D'Errico}
\affiliation{Nexus for Quantum Technologies, University of Ottawa, Ottawa, K1N 6N5, ON, Canada}
%%%%
\author{Hugo Larocque}
\affiliation{Research Laboratory of Electronics, Massachusetts Institute of Technology, Cambridge, MA 02139, USA}
%
%%%%%
\author{Eliahu Cohen}
\affiliation{Faculty of Engineering and the Institute of Nanotechnology and Advanced Materials, Bar-Ilan University, Ramat Gan, Israel}

%%%%%
\author{Ebrahim Karimi}
\affiliation{Nexus for Quantum Technologies, University of Ottawa, Ottawa, K1N 6N5, ON, Canada}

\begin{abstract}
Low-dimensional topological objects, such as knots and braids, have become prevalent in multiple areas of physics, such as fluid dynamics, optics and quantum information processing. Such objects also now play a role in cryptography, where a framed knot can store encoded information using its braid representation for communications purposes. The greater resilience of low-dimensional topological elements under deformations allows them to be employed as a reliable framework for information exchange. Here, we introduce a challenge-response protocol as an application of this construction for authentication. We provide illustrative examples of both procedures showing how framed links and braids may help to enhance secure communication.
\end{abstract}

\keywords{Framed knot, Secure communication}
\maketitle

\section{Introduction}\label{sec1}
\begin{figure*}[t]
\centering
\includegraphics[width=0.9\textwidth]{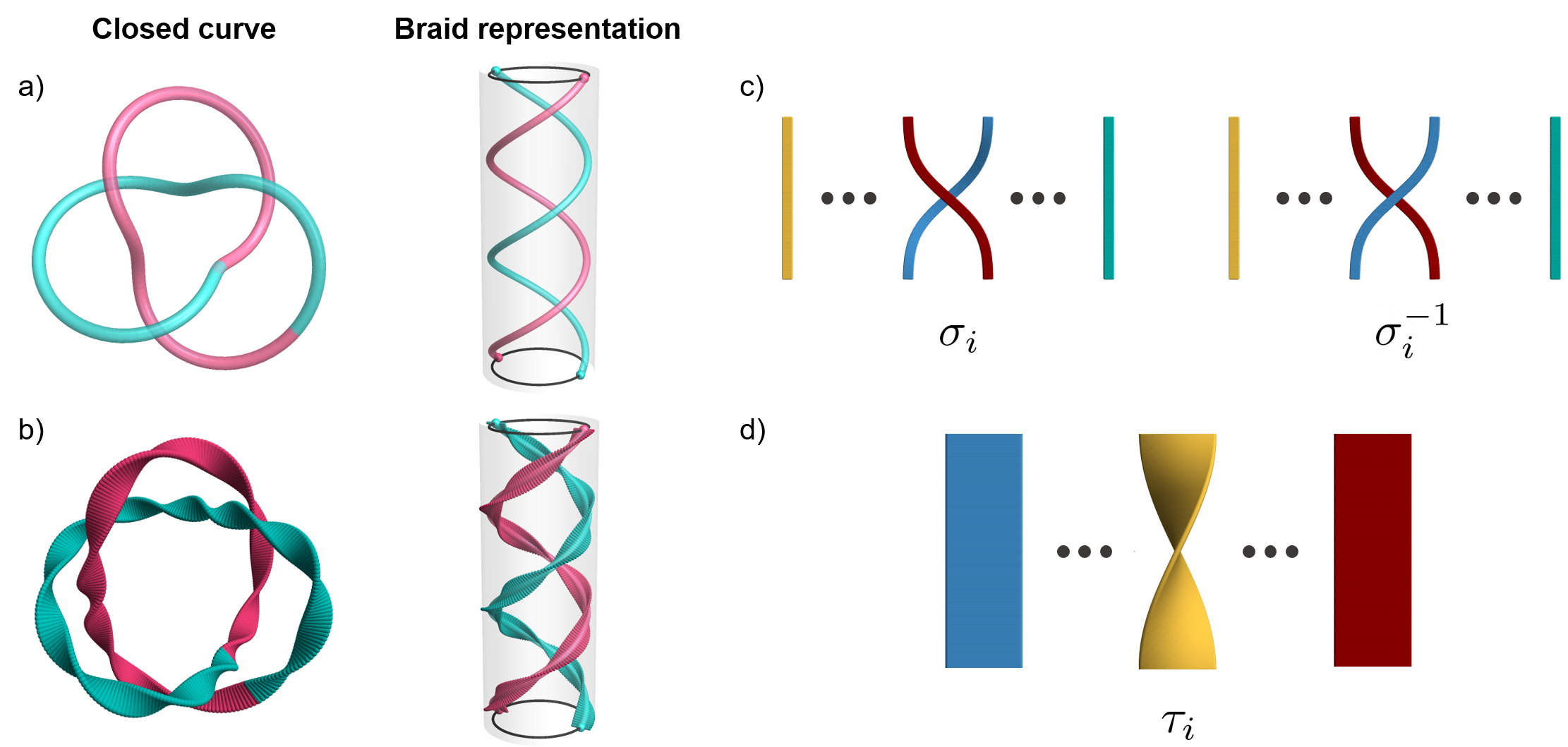}
\caption{\label{fig:fig1}\textbf{Brief description of braids and knots.} \textbf{a.} A classical knot can be understood as a possible configuration of a tangled string in space. The closure of a braid representation can yield any classical knot. \textbf{b.} A framed knot $(K,V)$ is visualized as a knotted ribbon with an even number of twists, and thus, framed knots are always orientable surfaces. \textbf{c.} The set of braid representations with $s$ strands forms the Artin braid group $\mathcal{B}_s$ under the braiding operator $\sigma_i$~\cite{prasolov1997knots}, which indicates the position exchange of the $i$-th and $(i+1)$-th strands in the braid. \textbf{d.} For a framed-knot, it is necessary to define another operator, $\tau_i$, to represent a single half-twist in the $i$-th strand.}
\end{figure*}

Alexandre-Th\'eophile Vandermonde was intrigued by knots and discussed their topological features in 1771. Later, in 1833, Carl Friedrich Gauss showed that the numerical invariant called today the linking number (representing the number of times that each curve winds around the other) can be calculated by an integral. Lord Kelvin's~\cite{kelvin1867vortex} attempt to develop a theory of fluids and solids based on microscopic vortex knots further stimulated the efforts to develop a rigorous mathematical theory of knots, links and braids. After Kelvin's atomic theory was abandoned, the research on knots developed separately from physics as a purely mathematical subject, with remarkable achievements as topology was extensively developed in the 20th century \cite{atiyah1995quantum}. Nowadays, knot theory has already reappeared in several areas of physics. Knot theory offers novel calculating tools, e.g. for exactly evaluating the partition function in vertex models \cite{akutsu1988knots} of statistical physics, in quantum field theory, especially concerning the study of Chern-Simons actions, of interest in both high energy and condensed matter physics~\cite{witten1989quantum}, in topological fluid dynamics~\cite{arnold1992topological}, for instance, in the study of energy minimization of linked vortex tubes~\cite{moffatt2021some}, in quantum information theory, particularly concerning topological quantum computation~\cite{bonesteel2005braid,stern2013topological,lahtinen2017short}, in electromagnetism, where knotted solutions of the Maxwell's equations have been discovered~\cite{ranada1990knotted,hoyos2015new,valverde2020numerical}, and in optics \cite{dennis2010isolated,leach2005vortex}. Knots and braid structures can be observed in a large variety of physical systems. They emerge from vortex lines in fluids, including cholesteric liquid crystals~\cite{sevc2014topological} and Bose-Einstein condensates~\cite{hall2016tying}, or from the interference of acoustic~\cite{sugic2020knotted,muelas2022observation} or optical waves in three-dimensions~\cite{ferrer2021polychromatic}, or when appropriate boundary conditions are imposed on paraxial light beams~\cite{dennis2010isolated, larocque2018reconstructing}. Knots and braidings can be observed or engineered in complex molecular structures, as proteins or DNA chains~\cite{lim2015molecular}.\\
The topological nature of these structures hints at their use for robust information distribution and manipulation. Candidate applications range from the above-mentioned topological quantum computing with non-Abelian anyons, to implementation of quantum money \cite{farhi2012quantum}, to alternative encodings of classical information. We proposed the latter in Ref.~\cite{larocque2020optical}, where we showed how specific optical paraxial beams are physical realizations of knotted ribbons (or framed knots). On such structures, numbers can be associated, not only with braiding operations, but also with the ribbon's twisting. From this representation, we proposed a prime encoding scheme. Alternative physical implementations based on nested knots generated from polychromatic optical fields have been proposed~\cite{kong2022high}. Here, we review the prime encoding in knotted ribbons scheme in more detail and generalize it to a full ``RSA-like'' key-distribution protocol.  \\
Geometrically, it is possible to describe a braid in $s$-strands, or $s$-braid, as a collection of $s$ non-intersecting strings joining two parallel planes. In physics, these low-dimensional topological elements have become relevant in the context of anyon-based topological quantum computing~\cite{nayak2008non}. While braids are fascinating mathematical objects by themselves, we focus our interest on their connection with knots and links. Figure~\ref{fig:fig1} illustrates that by tying the endpoints of the strands in a braid, one or more closed-knotted curves are obtained. Inversely, Alexander's braiding theorem states that every knot or link can be represented by the closure of a braid~\cite{alexander1923lemma}. It must be noted that the closure mapping is surjective: a knot may admit multiple different -- although equivalent -- braid representations.
A framed knot is an extension of classical knots, which arise from equipping the curves with a vector field normal to the curve at any point. This vector field is allowed to twist -- 2$\pi$-rotations -- around the curve~\cite{elhamdadi2020framed}. Therefore, we can visualize a framed knot as a tangled ribbon whose ends have been glued together after being twisted several times. As in the previous case, i.e. knots, it is possible to construct a framed braid representation for any framed knot or link. However, we encounter an irreducible arbitrariness in the construction of the braid representation. There is more than one way to specify a framing for a braid from which a particular knotted ribbon is formed.

In this Article, we introduce the use of low-dimensional elements -- framed knots and their corresponding braid representation -- as a reliable platform for secure communications. Low-dimensional elements exhibit greater resilience to environmental disturbances than conventional data structures~\cite{nazanin:22}. Therefore, the information content -- the myriad of topological invariants -- is generally insensitive to many deformations occurring when topological objects are materialized and handled. We further develop the encoding of prime numbers using the braid representation and the half-twists per strand. The privacy of our message is achieved when the knot is closed. Therefore, our protocol relies not only on the difficult endeavor of prime factorization but also on the non-unique framed braid representation admitted by a knot.

\section{Encoding prime numbers in a framed braid}\label{sec2}
A framed knot $(K,V)$ is obtained when we equip a knot $K$ with a continuous non-vanishing vector field $V$ normal to the curve at all points, referred to as \textit{framing}. Thus, it is possible to define an equivalent class for a framed knot using the framing integer $M$, i.e. the total number of half-twists. Since a framed knot $(K,V)$ may admit multiple braid representations, the braid representation is not unique, and the closure of a framed braid may act as an operation to encode information. Nevertheless, the message can be recovered only if the receiver is able to reconstruct the correct braid representation.

Let us start by considering a framed $s$-braid representation, in which each strand is identified by an index $k$. We can define $d_k$ as the number of half-twists along the $k$-th strand exhibiting half-twists. If we assign a \textit{distinct} prime number $p_k$ to each strand in our braid, then the framed $s$-braid as a whole captures the prime factorization of a natural number,
\begin{equation}
\label{eq:nat1}
    N=\prod_{\{k \mid d_k \neq -\infty\}} p_k^{(\alpha^{d_k})},
\end{equation}
where $\alpha$ is a real number chosen arbitrarily, (we point out that $\alpha$ is not related to any feature of the knot), and we assign $d_k=-\infty$ for untwisted strands. Note that the framing integer of the underlying framed knot obtained by the closure of the braid is given by $M = \sum_{\{k \mid d_k \neq -\infty\}} d_k$. In addition, we introduce a second quantity, which will be needed to construct the topological invariant of the braid representation,
\begin{equation}
\label{eq:beta}
    \beta = \prod_{\{k \mid d_k \neq -\infty\}} p_k^{\left(\alpha^{d_k - M}\right)},
\end{equation}
which permits us to write the number of twists in the braid representation as 
\begin{equation}
\label{eq:twists}
    M =\log_\alpha \left( \sum_{\{k \mid d_k \neq -\infty\}} \alpha^{d_k}\log_{\beta} p_k \right).
\end{equation}
Using these expressions, Eqs.~\eqref{eq:nat1},  \eqref{eq:beta} and \eqref{eq:twists}, we can rewrite the natural number $N$ as 
\begin{equation}
\label{eq:nat2}
N_{\alpha,\beta}(M) = \beta^{(\alpha^M)}. 
\end{equation}
In the latter expression, it becomes evident that $N$ is determined exclusively by the total number of twists in the framed knot, $M$, and the pair $(\alpha,\beta)$. Therefore, $N_{\alpha,\beta}(M)$ is a topological invariant of the knotted ribbon. By comparing Eq.~\eqref{eq:nat1} and Eq.~\eqref{eq:nat2}, it comes to light that the prime factorization of $N_{\alpha,\beta}(M)$ encodes the number of half-twists per strand in the frame braid representation of the knotted ribbon. 
\begin{figure}[!]
\centering
\includegraphics[width=0.5\textwidth]{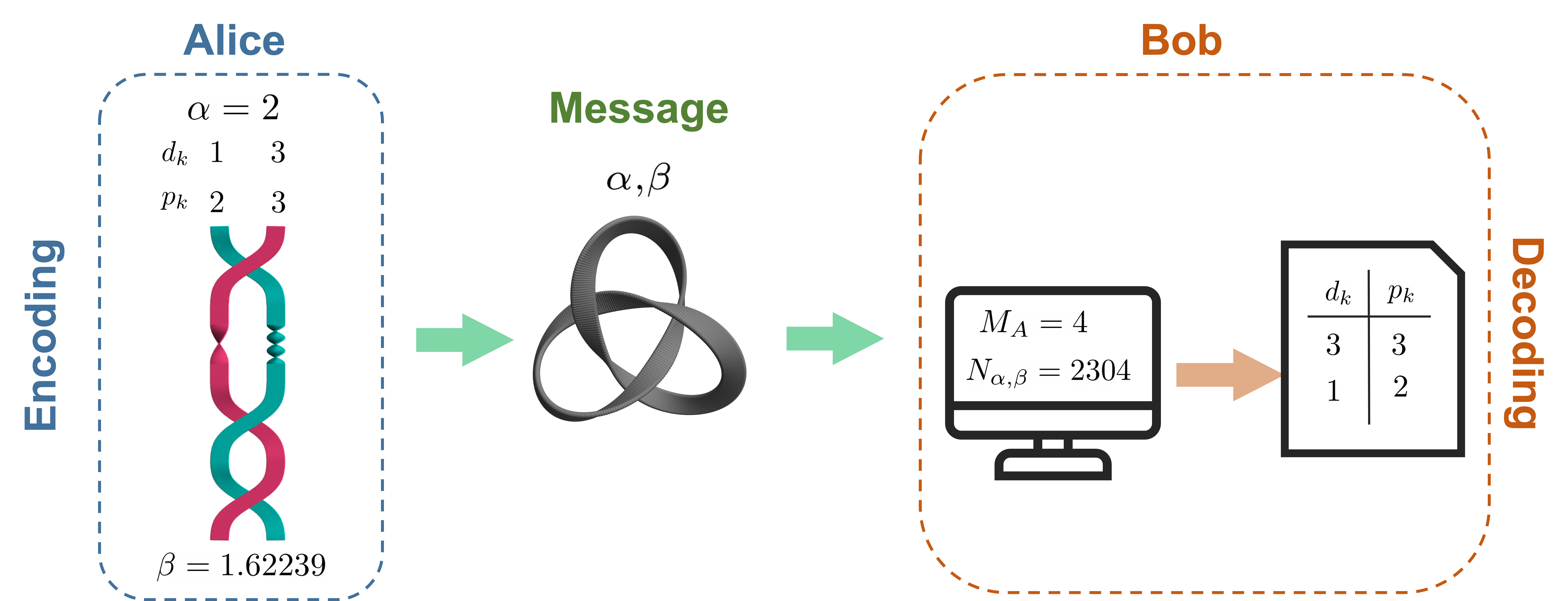}
\caption{\label{fig:fig2}\textbf{Encoding mechanism of prime numbers in framed braids.} Alice generates both the set of twists per strand and the prime numbers associated with each strand. After choosing $\alpha$, she computes $\beta$ and closes her braid, obtaining the knot $K_A$. Then, she sends her message composed of the pair of numbers $(\alpha,\beta)$ alongside the framed knot. Then Bob proceeds to calculate the natural number $N_{\alpha,\beta}(M)$ and computes its prime factorization. The number of half-twists per strand can be obtained by employing Eq.~\eqref{eq:nat1}.}
\end{figure}
\section{Sharing information using knotted ribbons}\label{sec3}
Given the relationship between the braid representation of a framed knot and the prime factorization, we propose the following scheme to share information \textit{securely}. A protocol sketch is depicted in Fig.~\ref{fig:fig2}. Let us consider a scenario where Alice decides to send a message to Bob. The message is computed from initial inputs, i.e., a set of integers $d_k$ where $k=1,\, 2, \dots, s$. In our case, Alice's message is encoded using a framed $s$-braid such that $\{d_k\}$ represent the half-twist number per strand. The procedure followed by Alice, illustrated in Fig.~\ref{fig:fig2}, can be summarized as follows: 
\begin{enumerate}
    \item Alice determines the set of numbers $d_k$ that she wants to send to Bob. Therefore, the number of strands $s$ in her braid representation is given by the length of $\{d_k\}$. In our example, the corresponding integers are $d_1=1$ and $d_2=3$, , leading to the total number of half-twists $M=4$.
    \item Alice allocates different prime numbers $p_k$ to each of the strands in the braid. For the sake of simplicity, we have used the first two prime numbers $p_1=2$ and $p_2=3$.  
    \item Then, she applies multiple braiding operators and closes the framed braid to obtain the knotted ribbon $K_A$. In this case, we opted for a trefoil knot. 
    \item At the same time, she chooses a positive number $\alpha$ and computes $\beta$ using Eq.~\eqref{eq:beta}. In our example, we set $\alpha=2$, yielding to a $\beta=2^{2^{-3}}\cdot 3^{2^{-1}} \approx 1.62239$.
\end{enumerate}
Then, Alice sends a package containing the $K_A$ and the pair $(\alpha,\beta)$ to Bob. Upon receiving these, Bob computes $N_{\alpha,\beta}(M)$ using Eq.~\eqref{eq:nat2} and its prime factorization. From here, Bob can retrieve $\{d_k\}$ and reconstruct an equivalent framed $s$-braid to the one sent by Alice. In our example, $N_{\alpha,\beta}(M)=2304$ and its prime factorization yields $2304=2^8 \cdot 3^2$.

Before continuing, it is useful to emphasize some subtleties in the protocol. First, it must be noted that the type of knot obtained from the closure of the generated braid has no impact on the encoding mechanism. However, the privacy of the encoding arises from the fact that a framed knot may admit multiple equivalent braid representations. If an eavesdropper intercepts the message, they should guess the number of braids and the way the half-twists are distributed among them. Alongside this, our encoding protocol exploits the difficulty of the prime factorization task, which has been a hard problem to solve on a classical computer. This implies that, when the prime numbers are sufficiently large, there is no efficient general algorithm that can perform the computation~\cite{arora2009computational}.

\section{Challenge-response protocol using framed links}\label{sec5}
\begin{figure}[!]%
\centering
\includegraphics[width=0.5\textwidth]{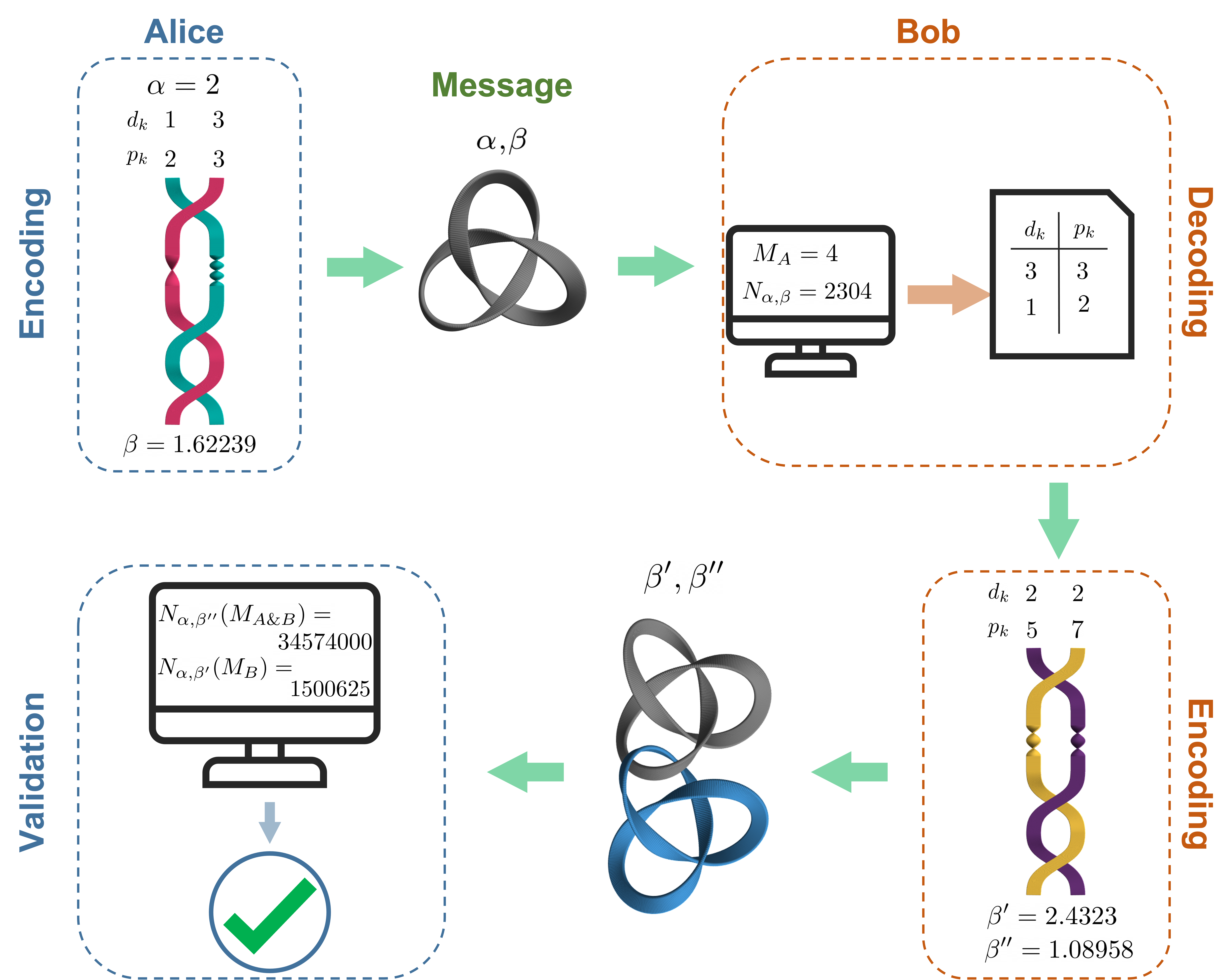}
\caption{\textbf{Challenge-response algorithm.} After Bob requests access to a service, Alice sends a message that contains a framed knot $K_A$ and the pair of numbers $(\alpha,\beta)$. Upon reception, Bob reconstructs Alice's braid representation. From here, he builds his own braid representation and makes sure that he assigns prime factors different from the ones sent by Alice. Before closing his braid, he links his knot $K_B$ to $K_A$, generating the link $K_A\&K_B$. Subsequently, he generates the message containing the link $K_A\&K_B$ and the pair $(\beta',\beta'')$, which are calculated using Eq.\eqref{eq:beta} from $K_B$ and $K_A\&K_B$, respectively. Alice calculates the topological invariant $N_{\alpha,\beta''}(M_{A\&B})$ for the link and $N_B$ and verifies if Bob's answer is correct using Eq.\eqref{eq:coprimes}.}\label{fig:fig3}
\end{figure}

Before continuing with an additional application, let us explore some properties of the topological invariant $N_{\alpha,\beta}(M)$ in the context of linked knots. We consider two framed knots, $K_A$ and $K_B$, whose total number of half-twists is $M_A$ and $M_B$, respectively. Then, suppose we connect the two knots forming the framed link $K_A\&K_B$ (See Figure \ref{fig:fig3}). A natural question arises at this stage: how does the topological invariant $N_{\alpha,\beta''}(M_{A\&B})$ of the link relates to the ones from the individual knots $N_{\alpha,\beta}(M_{A})$ and $N_{\alpha,\beta'}(M_{B})$? It can be readily verified using Eq.~\eqref{eq:nat1} that the number associated with the framed link can be written as 
\begin{equation}
\label{eq:coprimes}
N_{\alpha,\beta''}(M_{A\&B}) = N_{\alpha,\beta}(M_{A}) N_{\alpha,\beta'}(M_{B})
\end{equation} 
as long $N_{\alpha,\beta}(M_{A})$ and $N_{\alpha,\beta'}(M_{B})$ are co-prime. This is possible if and only if a different prime factor is assigned to each strand in the individual knots forming the link. 

In this spirit, we assume that both Alice and Bob can generate and reconstruct framed knots. Now, Bob is trying to access a service managed by Alice. For security reasons, Alice challenges Bob to perform some non-trivial calculations. If Bob's response is valid, Alice grants access to him. For that reason, the challenge-response authentication protocol goes as follows.
\begin{itemize}
\item Alice generates and sends $K_A$ and the pair $(\alpha,\beta)$ to Bob. For the sake of simplicity, we assume she chooses a braid with 2 strands and assigns the first two prime numbers $p^A_1=2$ and $p^A_2=3$, correspondingly. In addition, she allocates $d^A_1=3$ and $d^A_2=1$ half-twists on the framed strands and closes the braid to create a trefoil knot $K_A$. Here, $\alpha=2$, while the computation of $\beta$ yields 1.62239. 
\item Subsequently, Bob computes $N_{\alpha,\beta}(M_A)=2304$ from which he can obtain the prime factors in Alice's framed braid, $p_k^A$, $k=1,2,\ldots$.
\item To generate his response knot $K_B$, Bob generates two new sets of numbers; the first one is a set of prime numbers $\{p_1^B=5,p_2^B=7\}$ that are different from the ones provided by Alice and follow the condition $p_j^B < N_{\alpha,\beta}(M_A)$, the second set of numbers corresponds to the number of twists per strand in his braid representation. Here, he chooses $d_1^B=d_2^B=2$.
\item Based on this, he generates his knot $K_B$ and computes the corresponding $\beta'=2.4323$. Before sending his answer to Alice, he creates the link $K_A\&K_B$ and calculates $\beta"=1.0895$. Finally, Bob's answer to the challenge is a package that includes $\beta',\, \beta"$ and the link $K_A\&K_B$.  
\end{itemize}

According to Eq.~\eqref{eq:coprimes}, Alice should be able to compute $N_{\alpha,\beta'}(M_{B})$ as $N_{\alpha,\beta''}(M_{A\&B})/N_{\alpha,\beta}(M_{A})$. To verify that this number is indeed co-prime to $N_{\alpha,\beta}(M_{A})$, she might attempt to divide it by any of her factors.  This approach may nevertheless be time-consuming for large integers (the time complexity for calculating the totient function $\varphi(x)$ is currently known to scale no better than $\sqrt{x}$). Alternatively, Bob could include an integer defined as
\[
\gamma=b^{\varphi(N_{\alpha,\beta'}(M_B))} \, \text{mod} \, N_{\alpha,\beta''}(M_{A \& B})
\]
%Definition of modulus 
where $b$ is co-prime to $N_{\alpha,\beta''}(M_{A \& B})$, $\varphi(x)$ is Euler's totient function and $\mod$ stands for the modulus function. Computing the totient function $\varphi(x)$ may be time consuming for a large $x$. Nevertheless, it is readily computed given $x$'s prime factors, i.e., $\varphi(x) = \prod_{k=1}^n (p_k^{j_k} -1 )$, where $x =\prod_{k=1}^n p_k^{j_k}$. This alludes to the role of Alice and Bob's totient functions, $\varphi(N_{\alpha,\beta}(M_A))$ and
$\varphi(N_{\alpha,\beta'}(M_B))$, as private keys in the above protocol. Indeed, these numbers are not exchanged, and they are kept on the respective participant sides throughout their interaction. Alice's quick check at the end of the protocol is based on two basic results in number theory: Fermat's little theorem states that 
\begin{equation*}
b^{\varphi(x)} = 1 \, \text{mod} \,x
\end{equation*}
for $b$ coprime to $x$. The totient function is a multiplicative
function, meaning that $\varphi(x\cdot y) = \varphi(x) \varphi(y)$ for $x$
coprime to $y$. Therefore, for any two coprime numbers $x$ and $y$
whose product is coprime to $b$,
\begin{equation*}
    \left(b^{\varphi(x)} \right)^{\varphi(y)} = 1  \, \text{mod} \, xy.
\end{equation*}

The methods described above allow Alice and Bob to maintain a secure information exchange using framed links. The authentication protocol complements and strengthens a natural extension of the previously introduced scheme of employing structured light in the form of a framed knot as information carriers \cite{larocque2020optical}. Employing two linked framed knots, referred to as a framed link, thus allows to establish an authentication protocol between two parties. Mathematically, the security behind the protocols lies at the heart of a mathematical task believed to be hard, which is similar in essence to Rivest-Shamir-Adleman (RSA) encryption.

\section{Conclusions}\label{sec6}
Our encoding protocols only utilize the number of twists in the framed braid and exclude its crossings. In conjunction with framed knots being able of representing multiple framed braids, this feature adds considerable flexibility in constructing the knot to be sent with the set of numbers $(N, \alpha, \beta)$. Hence, this resulting degree of freedom might prove to be essential in scenarios where the knot must be generated in a physical system bounded by certain laws, such as fluid dynamics~\cite{kleckner2013creation} or optical diffraction~\cite{dennis2010isolated,larocque2018reconstructing,larocque2020optical}, which might make generating some knots harder than others. The main contribution of this Article is a challenge-response authentication protocol based on framed links whose security relies on solving a classically hard computational problem. Optical vortex knots are indeed an interesting realization of these structures. Experimental realization of relatively simple knot structures (Hopf links, trefoils, cinquefoil, and figure-8 knots) have been reported \cite{larocque2018reconstructing, larocque2020optical}. Schemes for generating more complex optical knots based on the Milnor polynomial \cite{dennis2010isolated} have been theoretically proposed and used for the above-mentioned structures. Complex optical vortex knots may be more challenging to generate due to limited resolution of modern spatial light modulators, but an in-depth study in this direction has not been carried so far. Regardless of the specific realization, the topological nature of these structures suggests how they might be robust to a large class of imperfections that can be met in optical communication systems (imperfect optics, turbulence, finite apertures). The linked knot structures, necessary for the validation in the challenge-response protocol, may be generated by generalizing current techniques used for optical knots, for example encoding the two knots on different frequencies or polarisations and independently shifting their position in space in such a way to introduce a link in the singular skeleton.  The results of this work may help identifying what are the families of optical fields needed for knot-based secure communication.

\section{Acknowledgments} 
This work was supported by the High Throughput and Secure Networks Challenge Program at the National Research Council of Canada, the Canada Research Chairs (CRC), and Canada First Research Excellence Fund (CFREF) Program. This research was also supported by Grant No. FQXi-RFP-CPW-2006 from the Foundational Questions Institute and Fetzer Franklin Fund, a donor-advised fund of Silicon Valley Community Foundation. E.C. was supported by the Israeli Innovation Authority under Projects No. 70002 and No. 73795, by the Pazy Foundation, by the Israeli Ministry of Science and Technology, and by the Quantum Science and Technology Program of the Israeli Council of Higher Education.

\section*{Declarations}
The authors declare no competing financial interests. Correspondence and requests for materials should be addressed to A.C. (avcarmi@bgu.ac.il) or E.K. (ekarimi@uottawa.ca).

%\bibliographystyle{abbrv}
%\bibliography{References.bib}

\begin{thebibliography}{0}%
\makeatletter
\providecommand \@ifxundefined [1]{%
 \@ifx{#1\undefined}
}%
\providecommand \@ifnum [1]{%
 \ifnum #1\expandafter \@firstoftwo
 \else \expandafter \@secondoftwo
 \fi
}%
\providecommand \@ifx [1]{%
 \ifx #1\expandafter \@firstoftwo
 \else \expandafter \@secondoftwo
 \fi
}%
\providecommand \natexlab [1]{#1}%
\providecommand \enquote  [1]{``#1''}%
\providecommand \bibnamefont  [1]{#1}%
\providecommand \bibfnamefont [1]{#1}%
\providecommand \citenamefont [1]{#1}%
\providecommand \href@noop [0]{\@secondoftwo}%
\providecommand \href [0]{\begingroup \@sanitize@url \@href}%
\providecommand \@href[1]{\@@startlink{#1}\@@href}%
\providecommand \@@href[1]{\endgroup#1\@@endlink}%
\providecommand \@sanitize@url [0]{\catcode `\\12\catcode `\$12\catcode
  `\&12\catcode `\#12\catcode `\^12\catcode `\_12\catcode `\%12\relax}%
\providecommand \@@startlink[1]{}%
\providecommand \@@endlink[0]{}%
\providecommand \url  [0]{\begingroup\@sanitize@url \@url }%
\providecommand \@url [1]{\endgroup\@href {#1}{\urlprefix }}%
\providecommand \urlprefix  [0]{URL }%
\providecommand \Eprint [0]{\href }%
\providecommand \doibase [0]{http://dx.doi.org/}%
\providecommand \selectlanguage [0]{\@gobble}%
\providecommand \bibinfo  [0]{\@secondoftwo}%
\providecommand \bibfield  [0]{\@secondoftwo}%
\providecommand \translation [1]{[#1]}%
\providecommand \BibitemOpen [0]{}%
\providecommand \bibitemStop [0]{}%
\providecommand \bibitemNoStop [0]{.\EOS\space}%
\providecommand \EOS [0]{\spacefactor3000\relax}%
\providecommand \BibitemShut  [1]{\csname bibitem#1\endcsname}%
\let\auto@bib@innerbib\@empty
%</preamble>
\end{thebibliography}%


\begin{thebibliography}{10}


\bibitem{kelvin1867vortex}
L.~Kelvin.
\newblock On vortex atoms.
\newblock In {\em Proc. R. Soc. Edin}, volume~6, pages 94--105, 1867.

\bibitem{atiyah1995quantum}
M.~Atiyah.
\newblock Quantum physics and the topology of knots.
\newblock {\em Reviews of Modern Physics}, 67(4):977, 1995.

\bibitem{akutsu1988knots}
Y.~Akutsu and M.~Wadati.
\newblock Knots, links, braids and exactly solvable models in statistical
  mechanics.
\newblock {\em Communications in Mathematical Physics}, 117(2):243--259, 1988.

\bibitem{witten1989quantum}
E.~Witten.
\newblock Quantum field theory and the jones polynomial.
\newblock {\em Communications in Mathematical Physics}, 121(3):351--399, 1989.

\bibitem{arnold1992topological}
V.~I. Arnold.
\newblock Topological methods in hydrodynamics.
\newblock {\em Vladimir I. Arnold-Collected Works}, pages 433--454, 1992.

\bibitem{moffatt2021some}
H.~Moffatt.
\newblock Some topological aspects of fluid dynamics.
\newblock {\em Journal of Fluid Mechanics}, 914, 2021.

\bibitem{bonesteel2005braid}
N.~E. Bonesteel, L.~Hormozi, G.~Zikos, and S.~H. Simon.
\newblock Braid topologies for quantum computation.
\newblock {\em Physical Review Letters}, 95(14):140503, 2005.

\bibitem{stern2013topological}
A.~Stern and N.~H. Lindner.
\newblock Topological quantum computation--from basic concepts to first
  experiments.
\newblock {\em Science}, 339(6124):1179--1184, 2013.

\bibitem{lahtinen2017short}
V.~Lahtinen and J.~Pachos.
\newblock A short introduction to topological quantum computation.
\newblock {\em SciPost Physics}, 3(3):021, 2017.

\bibitem{ranada1990knotted}
A.~F. Ranada.
\newblock Knotted solutions of the Maxwell equations in vacuum.
\newblock {\em Journal of Physics A: Mathematical and General}, 23(16):L815,
  1990.

\bibitem{hoyos2015new}
C.~Hoyos, N.~Sircar, and J.~Sonnenschein.
\newblock New knotted solutions of maxwell's equations.
\newblock {\em Journal of Physics A: Mathematical and Theoretical},
  48(25):255204, 2015.
  
\bibitem{valverde2020numerical}
A.~M. Valverde, L.~D. Angulo, M.~R. Cabello, S.~G. Garc{\'\i}a, J.~J. Omiste,
  and J.~Luo.
\newblock Numerical simulation of knotted solutions for Maxwell equations.
\newblock {\em Physical Review E}, 101(6):063305, 2020.

\bibitem{dennis2010isolated}
M.~R. Dennis, R.~P. King, B.~Jack, K.~O\'holleran, and M.~J. Padgett.
\newblock Isolated optical vortex knots.
\newblock {\em Nature Physics}, 6(2):118--121, 2010.

\bibitem{leach2005vortex}
J.~Leach, M.~R. Dennis, J.~Courtial, and M.~J. Padgett.
\newblock Vortex knots in light.
\newblock {\em New Journal of Physics}, 7(1):55, 2005.

\bibitem{sevc2014topological}
D.~Se{\v{c}}, S.~{\v{C}}opar, and S.~{\v{Z}}umer.
\newblock Topological zoo of free-standing knots in confined chiral nematic
  fluids.
\newblock {\em Nature Communications}, 5(1):3057, 2014.

\bibitem{hall2016tying}
D.~S. Hall, M.~W. Ray, K.~Tiurev, E.~Ruokokoski, A.~H. Gheorghe, and
  M.~M{\"o}tt{\"o}nen.
\newblock Tying quantum knots.
\newblock {\em Nature Physics}, 12(5):478--483, 2016.

\bibitem{sugic2020knotted}
D.~Sugic, M.~R. Dennis, F.~Nori, and K.~Y. Bliokh.
\newblock Knotted polarizations and spin in three-dimensional polychromatic
  waves.
\newblock {\em Physical Review Research}, 2(4):042045, 2020.

\bibitem{muelas2022observation}
R.~D. Muelas-Hurtado, K.~Volke-Sep{\'u}lveda, J.~L. Ealo, F.~Nori, M.~A.
  Alonso, K.~Y. Bliokh, and E.~Brasselet.
\newblock Observation of polarization singularities and topological textures in
  sound waves.
\newblock {\em arXiv preprint arXiv:2210.03976}, 2022.

\bibitem{ferrer2021polychromatic}
M.~F. Ferrer-Garcia, A.~D'Errico, H.~Larocque, A.~Sit, and E.~Karimi.
\newblock Polychromatic electric field knots.
\newblock {\em Physical Review Research}, 3(3):033226, 2021.

\bibitem{larocque2018reconstructing}
H.~Larocque, D.~Sugic, D.~Mortimer, A.~J. Taylor, R.~Fickler, R.~W. Boyd, M.~R.
  Dennis, and E.~Karimi.
\newblock Reconstructing the topology of optical polarization knots.
\newblock {\em Nature Physics}, 14(11):1079--1082, 2018.

\bibitem{lim2015molecular}
N.~C. Lim and S.~E. Jackson.
\newblock Molecular knots in biology and chemistry.
\newblock {\em Journal of Physics: Condensed Matter}, 27(35):354101, 2015.


\bibitem{farhi2012quantum}
E.~Farhi, D.~Gosset, A.~Hassidim, A.~Lutomirski, and P.~Shor.
\newblock Quantum money from knots.
\newblock In {\em Proceedings of the 3rd Innovations in Theoretical Computer
  Science Conference}, pages 276--289, 2012.
  
\bibitem{larocque2020optical}
H.~Larocque, A.~D'Errico, M.~F. Ferrer-Garcia, A.~Carmi, E.~Cohen, and
  E.~Karimi.
\newblock Optical framed knots as information carriers.
\newblock {\em Nature Communications}, 11(1):5119, 2020.

\bibitem{kong2022high}
L.-J. Kong, W.~Zhang, P.~Li, X.~Guo, J.~Zhang, F.~Zhang, J.~Zhao, and X.~Zhang.
\newblock High capacity topological coding based on nested vortex knots and
  links.
\newblock {\em Nature Communications}, 13(1):2705, 2022.

\bibitem{nayak2008non}
C.~Nayak, S.~H. Simon, A.~Stern, M.~Freedman, and S.~D. Sarma.
\newblock Non-abelian anyons and topological quantum computation.
\newblock {\em Reviews of Modern Physics}, 80(3):1083, 2008.

\bibitem{alexander1923lemma}
J.~W. Alexander.
\newblock A lemma on systems of knotted curves.
\newblock {\em Proceedings of the National Academy of Sciences of the United
  States of America}, 9(3):93, 1923.

\bibitem{prasolov1997knots}
V.~V. Prasolov and A.~B. Sosinski{\u\i}.
\newblock {\em Knots, links, braids and 3-manifolds: an introduction to the new
  invariants in low-dimensional topology}.
\newblock Number 154. American Mathematical Soc., 1997.

\bibitem{elhamdadi2020framed}
M.~Elhamdadi, M.~Hajij, and K.~Istvan.
\newblock Framed knots.
\newblock {\em The Mathematical Intelligencer}, 42(4):7--22, 2020.

\bibitem{nazanin:22}
N.~Dehgan, A.~D'Errico, T.~Jaouni, and E.~Karimi.
\newblock Effects of aberration on vortex knots.
\newblock {\em in preparation}, 2022.

\bibitem{arora2009computational}
S.~Arora and B.~Barak.
\newblock {\em Computational complexity: a modern approach}.
\newblock Cambridge University Press, 2009.

\bibitem{kleckner2013creation}
D.~Kleckner and W.~Irvine.
\newblock Creation and dynamics of knotted vortices.
\newblock {\em Nature physics}, 9(4):253--258, 2013.


\end{thebibliography}

\end{document}